\documentclass[amsmath,amssymb,twocolumn,nofootinbib]{aastex6}
\newcommand\simlt{\lower.5ex\hbox{$\; \buildrel < \over \sim \;$}}
\newcommand\simgt{\lower.5ex\hbox{$\; \buildrel > \over \sim \;$}}

\begin{document}
\title{Do Galactic Antiprotons Come from Decaying Dark Matter?}
\author{David Eichler$^1$}

%\author{Noemie Globus$^{2}$}

\author{Raz Idan$^1$}

\author{Eyal Gavish$^1$}

\author{Tanguy Pierog$^{3}$}
%\author{Noemie Globus\altaffilmark{1,2} and David Eichler\altaffilmark{2}}
%\altaffiltext{1}{Racah Institute of Physics, The Hebrew University, 91904 Jerusalem, Israel}
%\altaffiltext{2}{Dept. of Physics, Ben-Gurion University, Be'er-Sheba 84105, Israel}
\affiliation{$^{1}$Physics Department, Ben-Gurion University, Be'er-Sheba 84105, Israel}
\affiliation{$^{2}$Racah Institute of Physics, The Hebrew University, 91904 Jerusalem, Israel}
\affiliation{$^{3}$Karlsruhe Institute of Technology, Postfach 3640, 76021 Karlsruhe, Germany}

\begin{abstract}
{  It is shown that the antiproton spectrum  recently reported by the  AMS02 collaboration can be accounted for by dark matter (DM)  decay if the residence time in the Galactic halo is of order 90 Myr. The DM lifetime assumed, $ 5 \cdot 10^{27}$ s  for DM particle masss of 3 TeV, is shown to be consistent with the constraints of diffuse gamma ray background. Alternative sources of antiprotons, such as those from compact objects,  are shown to be strongly constrained by neutrino astronomy and therefore unlikely candidates.}
\end{abstract}

\section{Introduction}

{  The AMS02 collaboration has recently reported an antiproton  component in Galactic cosmic rays (Agulilar et al, 2016) that has the same spectrum as the primary protons. This would be surprising if they were merely secondaries from collisions in the interstellar medium, because other  secondaries from spallation  in the interstellar medium have steeper spectra than  primaries. The possibility should therefore be considered that there is some other source of antiprotons in the Galaxy other than collisions of cosmic ray protons in the interstellar medium.\footnote{ It has been claimed \citep{2017JCAP...02..048W} that the observed $\bar p$s are  inconsistent only  at the 2.1$\sigma$ level with what is expected from propagation models, but this does not eliminate the motivation for considering other sources of $\bar p$s.}   The two possible source classes that come to mind are a) dark matter decay in the disk and halo, and b) compact objects in the Galaxy. }

Dark nonbaryonic matter (DM) is widely believed to comprise most of the  matter density in the universe.
 Several forms of dark matter have been proposed, including weakly interacting massive particles (WIMPs), { light axions}, charged massive particles (CHAMPs), and massive astrophysical compact halo objects (MACHOs).  CHAMPs were strongly constrained by a variety of considerations, including sensitive rare isotope searches, and are seldom discussed anymore. Unclustered solar mass MACHOs have been strongly constrained by microlensing to comprise less than 10 percent of the Galactic dark matter, and wider mass ranges may also be constrained in the future  by pulsar timing \citep{2009ApJ...694L..95E} and by gravitational lensing searches for fast radio bursts \citep{2016PhRvL.117i1301M}. WIMPS, if they comprise the dark matter,  must have had a velocity-averaged annihilation rate constant in the early universe  of $<\sigma_{ann}|v|> \sim3 \cdot 10^{-26} \rm cm^{3} \,s^{-1}$ in order that the present mass density conform the astrophysical indications and this does not, under quite general assumptions,  seem to be enough to produce a detectable cosmic ray signal.  %The question has been raised as to whether their annihilation products are identifiable in the cosmic radiation \citep{1984PhRvL..53..624S}. 
%For WIMP masses of less than 300 GeV, this would imply an annihilation rate in galaxies that would impact astrophysical observations such as  high energy $\gamma$- ray emission from other galaxies, and failure to observe such consequence then constrains the WIMP mass to be larger than this. Large WIMP masses, on the other hand,  imply that their annihilation may be undetectable and unable to account for astrophysical anomalies.

%The WIMP density is only $ \sim 10^{-4}  (m_{DM}/3 \,\rm TeV)^{-1}  cm^{-3}$.  Assuming the thermal annihilation rate constant discussed above, the annihilation rate density, at this average DM particle density, would  be  only $10^{-33.5}(m_{DM}/ \rm 3TeV)^{-2} cm^{-3} s^{-1}$, which is insufficient.  The effective annihilation rate could be hypothesized to be larger by invoking concentrated clumps of dark matter, but simulations typically do not, {\it a priori}, predict enough of an enhancement.  Moreover, if such enhancement were due to clumpiness of the dark matter, then most of the prompt $\gamma$- rays that emerge from the annihilations would appear as hot spots in the high energy $\gamma$- ray background and would be detectable by the Fermi-Large Area Telescope and/or air Cerenkov arrays.  Other enhancement mechanisms, such as Sommerfeld enhancement due to some other attractive force that pulls dark matter particles together have been proposed, but they require an additional assumption.

Decaying WIMPs were also proposed as  possible contributors to the cosmic radiation  \citep[e.g.][]{1988PhLB..205..559B, 1989PhRvL..63.2440E}.  A specific mass range of several TeV, (hereafter TeV weakly unstable relic particles, or TWURPS), in any case a likely mass range for a number of reasons, was "predicted"  (\citet{1989PhRvL..63.2440E} and later in \citet{2009JCAP....01..043N}) by the numerical coincidence that  if  the {\it decay} is via interactions on the scale of Grand Unification (GUT), then for DM particle masses of $\sim 2$ TeV, the decay rate  could be just enough to contribute a detectable component to the Galactic cosmic radiation. %This coincidence was also noticed later  \citep{2009JCAP...01..043N}. 
 As the decay rate is proportional to $m_{DM}^5 \cos^4\theta$, where $\cos\theta$ is a mixing angle that establishes the coupling to any intermediate channel, the constraint of $m_{DM}$ is rather specific, given the GUT scale. Specifically, the decay rate is about $10^{-26.5}(10^{34} {\rm yr}/\tau_p)(m_{DM}/0.93\, {\rm TeV})^5 (\cos^4 \theta) \rm s^{-1}$ [where experimentally $ (10^{34} {\rm yr} /\tau_p) \lesssim 1$], giving a decay rate of   $\lesssim10^{-27} \rm s^{-1}$ in the Galaxy for a mass of $\gtrsim 1 $ TeV.

TWURPs remain both a candidate for  dark matter and for a source of detectable, very energetic anti-protons and $\gamma$-rays at detectable levels.  The specific interest in TWURPs is their cosmic ray signature,  and the purpose of this letter is to investigate whether a detectable antiproton signal is compatible with recently updated constraints on the $\gamma$-ray background.  

In contrast to an annihilation rate,  a decay rate does not increase with  density; so prior to the cosmic recombination era, less than $10^{-12}$ of the dark matter density would have been dissipated into heat, and this would have a negligible effect on the cosmic microwave background.  

As shown  in \citet{2016ApJ...822...56G},  many astrophysical extragalactic sources of energetic $\gamma$-rays yield softer spectra than the observed $\gamma$-ray background, especially at high redshift, where the $\gamma$-rays are subject to pair production, and the fit to the diffuse $\gamma$-ray background is improved when the relatively hard component from TWURP decay is added. However, accounting for the $\gamma$-ray background with yet unresolved astrophysical point sources  constrains the decay rate of dark matter \citep{2016ApJ...822...56G} and hence the flux of antiprotons ($\bar p$s) in the Galaxy that originate in such decay, and may provide an alternative source of hard $\gamma$-rays. 

{ In section II, the constraints on TWURP decay imposed by the $\gamma$-ray background are discussed. In section III, the allowed contribution to the Galactic $\bar p$ flux is displayed In section IV, it is argued that, the constraints on TWURP decay notwithstanding,   conventional (i.e. non-DM) astrophysical sources of $\bar p$s are unlikely to be as strong as the presently allowed signal from TWURP decay. (This is quite unlike the situation for positrons, for which conventional astrophysical sources are quite able to reproduce the anomalous high energy component reported by PAMELA).}  %Similarly, WIMPs that don't decay but merely annihilate have less of a chance of producing a detectable cosmic ray signature.  

\begin{figure}[!h]
\centering
\includegraphics[width=0.5\textwidth]{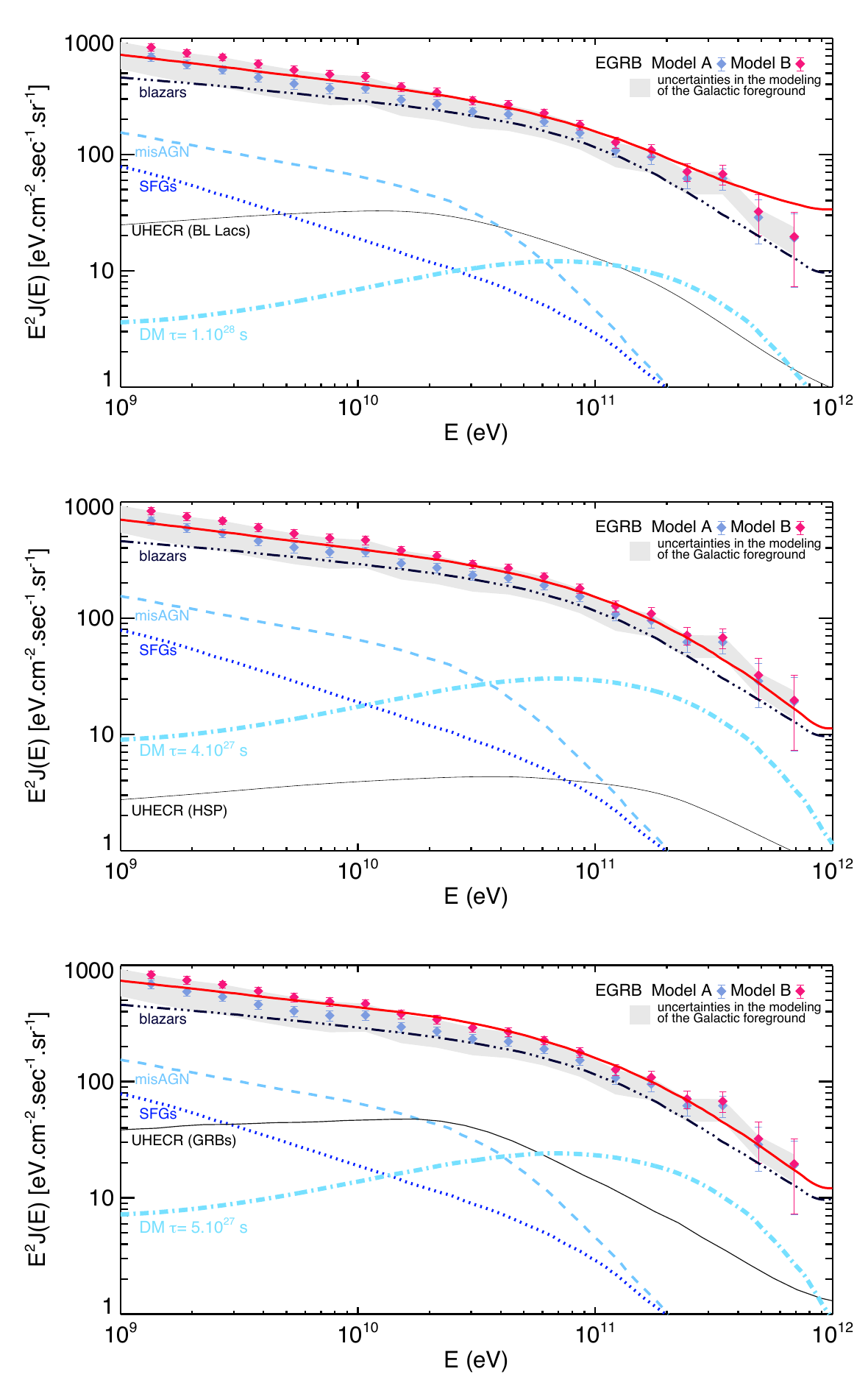}
\caption{  The $\gamma$-ray fluxes from extragalactic UHECR  sources, as calculated in \citep{2016ApJ...822...56G}, added to the contribution from TWURP decay and contributions from star forming galaxies (SFGs), misaligned AGNs (misAGN) and blazars, are plotted against the total extragalactic gamma-ray background (EGRB). Red solid line: total of all contributions. The contribution of the UHECRs is shown by the black solid line. The contribution of SFGs, taken from \citet{Ack2012}, is shown by the blue dotted line. The contribution of misaligned AGNs, taken from \citet{Inoue11}, is shown by the blue dashed line. The contribution of blazars (resolved and unresolved), taken from \citet{Ajello15}, is shown by the 3-dotted-dashed line. The contribution from TWURP decay is taken from \citep{2012JCAP...10..043M}, adjusted for different decay rates.
 {\it Upper panels:}  UHECR are assumed to be protons  from sources whose comoving density is independent of redshift 1+z. {\it Middle panel:} UHECR sources assumed to be protons from high synchrotron peak BL Lac objects evolving as $(1+z)^{-6}$ {\it Lower panel:} $\gamma$-rays originating from a mixed composition of UHECR originating from sources evolving as GRB \citep{2015MNRAS.451..751G} with a cosmological evolution of the source density  $\propto(1+z)^{2.1}$ for $z<3$ \citep{2010MNRAS.406.1944W}. In this model the proton contribution to the $\gamma$-ray flux is 60 percent, and the proton component undergoes a sharp cutoff (in $\exp^2$) above $E_{max,p}=10^{19.25}$ eV. %The DM lifetimes have been rescaled to suit the fit. 
 %The new estimate for the IGRB (pink shaded region around the red dots) put a new constraint on the DM lifetime of $\tau\ge 5\cdot10^{27}$ s.  
 The DM signal for a lifetime of $\tau= 4\cdot10^{27}$ s is the smallest lifetime allowed by the EGRB.}%, e.g. if UHECR sources are Galactic or extragalactic with a negative evolution scenario.}
\label{updated}
\end{figure}

\section{Limits on DM Decay from the Diffuse Gamma Ray Background}

 Recently, %limits on TWURP decay were obtained using recent data on the unresolved diffuse high energy $\gamma$-ray background and assuming that ultrahigh energy cosmic rays (UHECR) are pure protons \citep{2016ApJ...822...56G}.  TWURP decay,  particularly decay into the $W^+ -  W^- $ channel, can produce hard $\gamma $-rays, which  complement the $\gamma$-ray spectra predicted by several conventional astrophysical models of UHECR,  which are typically softer (because of the cosmic opacity introduced by extragalactic background light to $\gamma$-rays of several hundred GeV or more)  than the observed spectrum, which shows little if any cutoff effects at such high energy. %This is one motivation for specifically considering decay into the $W^+ - W^-$ channel.  Because the $W^+ - W^-$ channel produces harder $\bar p$ spectra than most other decay channels, it yields the $\bar p$ spectrum that is most likely to stand out against the background of secondary $\bar p$s from cosmic ray collisions.
it was found \citep{2016ApJ...822...56G} that the maximum TWURP decay rate  into the $W^+ - W^-$ channel that is consistent with the $\gamma$-ray data (for $m_{DM}= 3$ TeV) is $[3 \cdot 10^{27} \rm s]^{-1}$, { given any other assumptions about UHECRs}. The $W^+ -  W^-$ channel is consistently used here by way of example; however,  raising the TWURP mass makes the decay products harder in similar proportions, so if  $m_{DM}\simeq 3$ TeV is considered a flexible parameter then the constraints imposed by the $\gamma$-ray background on $\bar p$s should not depend dramatically on channel.
{ Moreover, the $W^+ - W^-$ channel produces harder $\bar p$ spectra than most other decay channels, it yields the $\bar p$ spectrum that is most likely to stand out against the background of secondary $\bar p$s from cosmic ray collisions.}

More recently, the Fermi  Collaboration has claimed that $86^{+16}_{-14}$\% of the total $\gamma$-ray background { (EGRB)} above 50 GeV can be attributed to BL Lac objects \citep{2016PhRvL.116o1105A}. %(though this fraction is likely to be energy dependent and in any case somewhat uncertain)
{{This fraction is calculated using the more constraining model for the Galactic foreground, namely Model A of \citet{2015ApJ...799...86A}. Let us note that a more careful analysis of the interpretation of the different components of the total $\gamma$-ray background still needs to be done before  the fraction of 86\% is firmly established. In fact, when considering Model B for the Galactic foreground \citep{2015ApJ...799...86A}, this fraction drops to $\sim$ $71^{+13}_{-12}$\%} \citep[][]{arXiv:1703.04158}}. %However, this leaves little room for UHECR protons at high redshift, which would produce $\gamma$-rays from UHECR-initiated pair production, and even UHECR proton sources with a constant comoving density are only marginally consistent with the $\gamma$-ray background {\bf when considering Model A}.  
{However, this leaves little room for UHECR protons at high redshift, which would produce $\gamma$-rays from UHECR-initiated pair production. In fact, the electron-positron dip (pure-proton) scenario rules out SFR-like and stronger cosmological evolutions \citep{2016ApJ...822...56G}. }
The implication is that if UHECR sources evolve as $\gamma$-ray bursts (GRB) or star formation (SFR), which were more common in the past, then they must either a) have a significant non-protonic component  or b) { have a substantial} Galactic { contribution} in the energy range that would otherwise dominate the contribution to the diffuse $\gamma$-ray background.  %We then a) use the limits on TWURP lifetime set by recent tightening of the limits imposed on the diffuse extragalactic $\gamma$-ray background to  set limits on the Galactic $\bar p$s contribution that could be detected by AMS02 and b) consider other possible astrophysical sources of high energy $\bar p$s and show that their plausibility is questionable.} 

 On the other hand, if the diffuse $\gamma$-ray contribution from UHECRs is due to ``present-biased'' evolution (i.e. comoving luminosity proportional to  $(1+z)^m$, $m\le  0$), such as BL Lac objects, then there is less secondary $\gamma$-ray emission from past UHECR production.  Moreover, the $\gamma$-ray spectra resulting from $m \le 0$ evolution are harder because there is less pair production opacity for $\gamma$-rays at hundreds of GeV,  so there is less of a ``need''  for  decaying TWURPs and  for the attendant $\bar p$s. %{\bf The middle and lower panels of figure \ref{figure1}} shows a fit to the {IGRB  } and the contribution of secondary $\gamma$-rays from the UHECR under the assumption that the UHECR come from a) sources whose comoving space density is constant in time (i.e. non-evolving) and b) sources that evolve as high synchrotron-peaked (HSP) BL Lac objects (comoving density proportional to $(1+z)^{-6}$, see \citep{2014ApJ...780...73A}). While TWURP decay into $W^+ - W^-$ still improves the fit, it plays a smaller role and makes a weaker case for such decay.  
  
% In light of the most recent claim by the Fermi collaboration that as much as 86  (+16,-14)\%  of the extragalactic $\gamma$-ray background  can be attributed to point sources, we have updated fits from \citep{2016ApJ...822...56G}, %Figure \ref{updated} 
%assuming a diffuse $\gamma$-ray background equal to the implied remainder, which we have taken to be a constant 14\% of the total above 50 GeV.  At the 1 sigma confidence level, the uncertainty in the background is probably about a factor of 2, and this factor can vary with energy. 

 { In  Figure \ref{updated}  we show fits to the total extragalactic $\gamma$-ray background  (EGRB), adding all the contributions (we took the estimation of the blazars contribution from  \citet{Ajello15}, misaligned AGN from \citet{Inoue11}, and star forming galaxies from  \citet{Ack2012}). The upper  panels corresponds to non-evolving UHECR protons sources. The middle panel corresponds to proton sources with negative evolution. The two models for the Galactic foreground (models A and B) are shown.
We see that, for proton sources, there is no room for a dark matter component in the non evolving scenario (a TWURP lifetime of $1\cdot 10^{28}$s was assumed in the upper panel). For a negative evolution we find a lower limit for the TWURP lifetime of $4\cdot 10^{27}$s.}
 However the pure proton scenario is a limiting case. Assuming a mixed composition, in the light of the recent measurements made by Auger \citep{Aab:2014aea}, one can fit the cosmic ray spectrum down to the ankle (which is in this case the signature of the end of the transition from Galactic to extragalactic cosmic rays) with a harder spectrum at the source (approximately in $E^{-2}$ for protons), so a mixed composition model predicts $\sim$ two times less   $\gamma$-ray background than a pure proton "dip" model %, which requires a spectral index of $\sim2.5$ for SFR evolution ($\sim2.4$ for GRB) in order to fit the ankle 
 \citep{2011A&A...535A..66D,2015PhRvD..92b1302G, 2016ApJ...822...56G}. The lower panel of Figure \ref{updated} shows the fits for the mixed composition scenario, { taken from \citet{2015MNRAS.451..751G}}.

% {\bf Moreover, assuming a mixed composition, in the light of the recent measurements made by Auger \citep{Aab:2014aea}, one can fit the cosmic ray spectrum down to the ankle (which is in this case the signature of the end of the transition from Galactic to extragalactic cosmic rays) with a harder spectrum at the source (approximately in $E^{-2}$), so a mixed composition model predicts $\sim 67 \%$ less   $\gamma$-ray background than a pure proton "dip" model, which requires a spectral index of $\sim2.5$ for SFR evolution ($\sim2.4$ for GRB) in order to fit the ankle \citep{2011A&A...535A..66D,2015PhRvD..92b1302G, 2016ApJ...822...56G}.
 Altogether, a lower limit on the TWURP lifetime (at $m_{DM} =3$ TeV) is emerging. %, and we favor this value to estimate the upper limit on the Galactic $\bar p$ flux. 
{  Allowing for Galactic foreground uncertainties,  the smallest allowable lifetime seems to be $4 \cdot 10^{27}$ s.   }

%Figure \ref{fig:DM_constraints} shows that the constraints imposed on the TWURP decay rate by considerations of the IGRB are stronger than those presently imposed   \citep{2015arXiv150707001D}, \citep{2015JCAP...09..023G}   by measurements of $\bar p$s in the Galactic CRs.

 \section{Estimating the $\bar p/p$ ratio from DM decay and comparison with other scenarios}
  The TWURP contribution scales with residence time and inversely with TWURP lifetime. Because the TWURP decay gives a harder component than secondary products of CR collisions, it can give a harder overall $\bar p/p$ ratio in the 100 - 300 GeV range. { In computing the $\bar p$ production rate density  from TWURP decay, we assumed that the local rest energy density of TWURPs in the Galactic disk is $0.3 \,\rm GeV\,  cm^{-3}$.  The $\bar p$ production spectrum was taken  from the PPPC data \citep{2011JCAP...03..051C}.  We simply assume that the local production rate density equals the local disappearance rate density, the latter being the density divided by the residence time.  The many uncertainties in the halo and propagation parameters are absorbed into this definition of the residence time.}
  The secondary $\bar p$ production rate we have assumed is taken from reference \citep{2015JCAP...09..023G}. 
	The secondary $\bar p/p$ ratio is  fixed by the relative spallation  and $\bar p$ production cross sections \citep{2014PhRvD..90h5017D}, and there is negligible uncertainty associated with the propagation.  However,  the  contribution from TWURP decay is  strongly dependent on the residence time in the halo, and, as stated above,  we regard this as very uncertain.\footnote{The Be$^{10}$ abundance sets a lower limit on this lifetime but not a model-independent upper limit.}

%\begin{comment}
%\begin{figure}
%
%
%\centering
%\includegraphics[width=0.5\textwidth]{Figure3.pdf}
%\caption{Lower limits on the DM lifetimes for a $DM \rightarrow W^+W^-$ decay as a function of the DM mass as obtained by \citep{2015arXiv150707001D} (black line) and by \citep{2015JCAP...09..023G} (red line). The vertical blue line is the possible values of lifetimes for the fits that do not include the expected blazar contribution to the  IGRB. The vertical magenta line is the DM lifetimes for the fits that include an anticipated  contribution from not-yet-resolved blazars to the $\gamma$-ray background. }
%\label{fig:DM_constraints}
%\end{figure}
%\end{comment}

\begin{figure}[!h]
\centering
\includegraphics[width=0.47\textwidth]{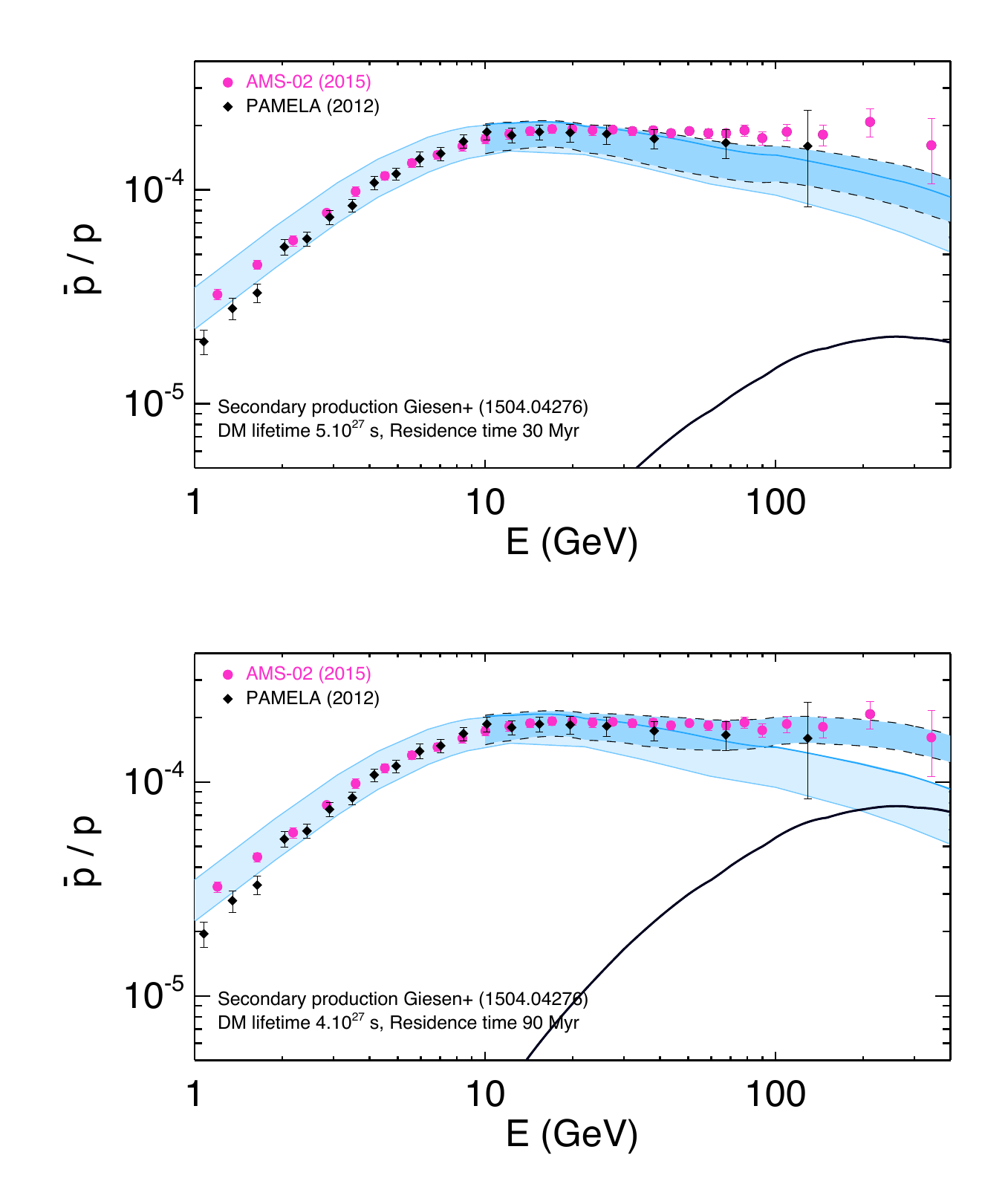}
\caption{The $\bar p/p$ ratio is plotted as a function of energy assuming a contribution from $p-p$ collisions with a collision probability that is consistent with the observed B/C ratio in the Galactic disk: the grammage traversed by a primary CR is assumed to be proportional to $E^{-\alpha}$ and a contribution from 3 TeV dark matter particles with a TWURP lifetime of $5\cdot 10^{27}$ s and for 
a residence time in the halo of $30(E/\rm 3\, GeV)^{-0.3}$ Myr. The secondary production is taken from \citep{2015JCAP...09..023G}.   The data point are from \citep{Adriani:2012paa} and \citep{PhysRevLett.117.091103}. The lower panel is for a TWURP lifetime of only $4\cdot 10^{27}$ s, the minimum compatible with  constraints from the diffuse $\gamma$-ray background, and a residence time of $90(E/\rm 3\, GeV)^{-0.3}$ Myr.}
\label{figbarp}
\end{figure}

\section{Alternative Astrophysical Sources of Antiprotons}
We have considered  whether alternate mechanisms for  "anomalous" $\bar p$ production (i.e. beyond what is anticipated from cosmic ray collisions in the interstellar medium),  e.g. from compact objects that  produce energetic hadrons \citep{1982Natur.295..391E}, could compete with dark matter decay in the 100-300 GeV energy range. We consider the following two scenarios:

Scenario a): A "sea'" of anomalous subrelativistic  $\bar p$s could exist in the Galaxy below 1 GeV.  Such $\bar p$s would be undetectable because they could not penetrate the solar system (without being modulated by the solar wind to beneath the range of detectability),  and some could be reaccelerated to beyond 100 GeV by supernovae remnant blast waves along with primary cosmic rays \citep{1980ApJ...237..809E}.

Scenario b):   Anomalous $\bar p$s could  be made at  energies above 100 GeV (e.g. by relativistic shocks associated with GRBs).  However, the data implies that  unanticipated $\bar p$s should {\it not} appear between 1 and 100 GeV, as observations already verify that  there is no observed excess there {beyond what is expected from cosmic ray collisions in the interstellar medium}.  Their sources must therefore { be harder than} the source of most cosmic rays, and  also need to be anomalously rich in $\bar p$s relative to CR protons  in order to cause an anomalous excess of the former but not of the latter. Such a scenario  has in fact been proposed \citep{2016ApJ...821L..24E} to explain UHECR.  { In this scenario, particles are accelerated in compact regions (e.g. GRB fireballs) where they undergo many collisions, and produce neutrons and antineutrons that escape the compact region, thereby avoiding much of the adiabatic loss that charged particles would suffer.   Because the antineutrons would decay into antiprotons, the question arises whether such a scenario  could yield   anomalous antiprotons at detectable levels. }

The maximum fraction $\epsilon_{\bar p} + \epsilon_{\bar n}$ 
of primary energy that can end up in $\bar p$s is roughly $\int [d\sigma_{\bar p}/dE_{\bar p}] E_{\bar p}dE_{\bar p}  /( E_{proj}\sigma_{inelastic})$, where $\sigma_{inelastic}$ is the total inelastic scattering cross section, $E_{proj}$ is the energy of the incident proton in the lab frame, and $d\sigma_{\bar p}/dE_{\bar p}$, the differential $\bar p$ production cross section, includes $\bar n$s. %From reference \citep{2011AdAst2011E..17J}, we estimate that this is about $ \epsilon_{\bar p} \epsilon_{\bar n}\simeq 1 \cdot 10^{-3}$.
We have run simulations using the EPOS LHC code \citep{Pierog:2013ria} and found that for primary energies of order 1 TeV,
  $\epsilon_{\bar p} \simeq 0.5 \cdot 10^{-2}$ and $\epsilon_{\bar n} \sim 1 \cdot 10^{-2}$.

As an observable level of anomalous $\bar p$s would suggest a  $\bar p$ luminosity $L_{\bar p}$ from our Galaxy of at least of order $10^{35}$ erg s$^{-1}$ (i.e. $10^{-4}$ of the CR luminosity above 100 GeV). In scenario a) the  actual required anomalous $\bar p$ luminosity from our Galaxy would then be about $10^{35}l / (f_r f_p)$ erg s$^{-1}$, where  $ l$ is the loss factor the $\bar p$s would undergo in being decelerated to below 1 GeV from above their production threshold of 10 GeV,  $f_r$ is the fraction of low energy $\bar p$s that get reaccelerated, and  $f_p$ is the fraction of energy, among the reaccelerated particles, in  particles that are reaccelerated to beyond 100 GeV. The loss factor $ l$ is  clearly at least 10.  The boron (B) to carbon (C) ratio, which decreases with energy over two decades of energy per nucleon, requires $f_r \ll 1$, because secondaries that are reaccelerated from low energy have the same spectrum as primaries \citep{ 1980ApJ...237..809E}, hence the B/C ratio would not decrease if the boron had a significant reaccelerated component. If the (presumably shock) reacceleration imparts a CR spectrum of $E^{-2-p}dE$, then $f_p \le 10^{-2p}$.  As the $\bar p$ production efficiency is at most $\sim 10^{-2}$, the primary cosmic rays that originally produced the $\bar p$s must have required at least $\sim 10^2 L_{\bar p}$.
Altogether, $10^{37}l / (f_r f_p)$ erg s$^{-1}$ are required in primary CR at $E \gtrsim 10^2$ GeV in order to produce the low energy $\bar p$s that would by hypothesis be reaccelerated to beyond 100 GeV. This would be comparable  to the total CR output from the Galaxy above $10^2$ GeV, and the question would arise as to why there is no evidence of this additional class of sources in the primary CR data.

Scenario b), in the version proposed by \citet{2016ApJ...821L..24E}, motivated by data above 1 EeV, does not produce $\bar p$s at the $10^{-4}$ level at AMS02-sensitive energies ($E\lesssim1$TeV).  %The spectrum of the primaries in the GRB must be close to $E^{-2}$, as opposed to $E^{-2.7}$ (and steeper beyond the knee) for Galactic CR protons, implying that if ex-neutrons dominate the proton flux at $E=3$ EeV, just below the ankle, then even if half are antineutrons, the antiproton flux resulting from antineutron decay at 100 GeV would be  below $10^{-5}$ of the proton flux. 
{ Similar scenarios, if designed to produce $\bar p$'s in the AMS02 sensitive region,   would produce too many UHE $\nu$s to be consistent with ICECUBE constraints. 
To see this, consider the %more general 
possibility that  Galactic GRBs release $\bar n$s at  $E\ll 1$ EeV.  }
If the $\bar p$s resulting from $\bar n$ decay are at $E\sim 300$ GeV, which is where an anomalous excess may have been detected by AMS02, then we need to consider that they underwent adiabatic losses by some factor $l$ after decaying, so that their energy at decay was $300\, l$ GeV.
If a Galactic GRB releases $10^{51}\epsilon_{51}$ ergs in neutrons of  $ 300E_{300} l m_n c^2$, then their decay length is about $1\cdot 10^{16}E_{300} l$ cm , and their decay volume is then $V=  (4\pi/3) 10^{48}E_{300}^3 l^3$ cm$^3$.  Their energy density is thus $U  = 1.6 \epsilon_{51,n} E_{300}^{-3}l^{-3} \cdot 10^{14} $ eV/cm$^3$. In order to relax to the interstellar energy density of $U_{is} \sim 1\, \rm eV/cm^3$,  the cloud of decayed neutrons must expand by a factor of  $l = [U/U_{is}]^{1/4} = [1.6 \cdot 10^{14}\epsilon_{51,n}E_{300}^{-3} l^{-3}]^{1/4}$, implying  
\begin{equation}
l =[1.6 \cdot 10^{14}\epsilon_{51,n}E_{300}^{-3}]^{1/7}.
\end{equation}

Thus we may estimate with good accuracy that    $l\sim 10^2 E_{300}^{-3/7}$, and that the $\bar n$  energy at decay is  $\sim3 \cdot 10^4 E_{300}^{4/7}$ GeV.  Hence, at $E_{\bar n} = 3 \cdot 10^4$ GeV,  { $l\simeq 10^2$, and  the $\bar n$ luminosity of compact sources  such as GRB within the  Galaxy  needed to account for the antiproton luminosity of the Galaxy $L_{\bar p} \simeq 10^{35}$ erg s$^{-1}$, would be $ L_{\bar p}l \simeq1 \cdot 10^{37}$ erg s$^{-1}$.}

But high energy $\nu$s  are produced with higher efficiency $\epsilon_{\nu}$ than $\bar n$s in $p-p$ collisions.  Using the EPOS LHC code \citep{Werner:2005jf,Pierog:2009cb,Pierog:2013ria}, we have estimated that about 20 percent of the primary energy $E_p$  goes into neutrinos, mostly in the energy range $0.05 \le E_{\nu} \le 0.1 E_p$, implying $\epsilon_{\nu}/\epsilon_{\bar n}\sim 20$.  For $E_p\sim 30 $ TeV, $E_{\nu}\gtrsim 5$ TeV,  and this is well within the range of sensitivity of ICECUBE. This would imply a neutrino luminosity from our Galaxy of $10^{37}\epsilon_{\nu}/\epsilon_{\bar n}$ erg s$^{-1}$, and, assuming it persists for $10^{17.5}$ s, the total amount of energy per unit mass produced is then $\sim 3 \cdot 10^{9} \epsilon_{\nu}/\epsilon_{\bar n}$ erg g$^{-1}$ $\gg 10^{10}$  erg g$^{-1}$.  On the other hand, the ICECUBE-detectable HE neutrino yield \citep{2015ApJ...805L...5A} implies a cosmic output of $ 2 \cdot 10^{-20}\rm erg \, cm^{-3}/\Omega_{DM}\rho_c  = 10^{10}$ erg g$^{-1}$, { where $\Omega_{DM}\rho_c$ is the cosmic dark matter density).}\footnote{unless the neutrinos have a sharp cutoff below 1 TeV and are hidden by the atmospheric background, but this would require unrealistically fine tuning.}   Moreover, less than $1.6 \cdot 10^{-2}$ of this is associated with (extragalactic) GRB \citep{2015ApJ...813L..10E}, suggesting that, if GRB from our Galaxy put out  a time average of  more than $10^{37.5} $erg s$^{-1}$ in UHE neutrinos, it would be highly exceptional.  

Note that {\it any}  steady Galactic source (or sources) that put out a time average of $10^{37}\epsilon_{\nu}/\epsilon_{\bar n}$ erg s$^{-1}$ in ICECUBE-sensitive neutrinos would have been readily detected by ICECUBE at the rate of $\sim 10^{2.5}\epsilon_{\nu}/\epsilon_{\bar n}$ neutrinos  per year, so we may in any case consider only episodic sources that occur in our Galaxy at a rate of  less than $\sim$1 per 10 yr,  so as to evade the monitoring that has been done so far by ICECUBE.  %(The exception is that the $\bar p$  source cut off below a TeV, so that the associated neutrinos are well below the sensitive range of ICECUBE.)  
In order to produce a time average of $10^{37}\epsilon_{\nu}/\epsilon_{\bar n}$ erg s$^{-1}\sim 2 \cdot 10^{38}$ erg s$^{-1}$ in neutrinos, they would have to produce more than  $10^{47.5} l$ ergs per occurrence in primary cosmic rays,  so that $\epsilon_{51}\gtrsim 10^{-3.5} l$.  This implies that  for {\it any} episodic Galactic source of detectable $\bar n$s, $l$ could not be much lower than $10^2$. Antiprotons  that are produced directly would presumably suffer even larger adiabatic losses.
 
We conclude that the alternative scenarios involving $\bar p$ production from $p-p$ collisions do not give efficient enough  $\bar p$ production and, at best, require fine tuning.  Thus observations of $\bar p$s by instruments such as AMS02 could provide evidence for or against the existence of slowly decaying TeV dark matter particles. Our analysis indicates that TWURP lifetimes of $10^{28}$ s and residence times in the Galactic halo of $10^{7.5}$ yr  give a $\bar p$ background that is at best marginally separable from the uncertainties in the secondary to primary ratio. A good fit to the recent AMS02 data is still achievable if UHECR are Galactic, or extragalactic with a mixed composition, allowing a TWURP decay time of $4 \cdot 10^{27}$s, and if there is a long $\bar p$  lifetime ($\sim 10^8$ yr) in the halo.% However, the uncertainties in the latter should decrease with time, as should the error bars of the $\bar p$ flux measurements.\\

 \section{Conclusions}

{ In conclusion, we have shown that the antiproton flux from dark matter, though strongly constrained by the EGRB, can account for the flat (i.e. energy independent) antiproton/proton ratio  in Galactic cosmic-rays above 100 GeV if the halo confinement time is of the order of 90 Myr.  Moreover, we have argued that less exotic astrophysical sources of collisional secondary cosmic-rays antiprotons are unlikely to be strong enough to account for a serious elevation of antiprotons at these energies.  We have tacitly assumed that the flat antiproton/proton ratio is surprising, because the fact that the B/C ratio decreases with energy is indicative of an escape rate from the Galaxy that increases with energy. If there is a different reason for the decreasing B/C ratio, then there would be less of a need for an exotic source of antiprotons.}\\

We gratefully acknowledge discussions with D. Allard,  R. Brustein,   G.  Beuf,  M. Cirelli, M. Lublinsky,  E. Parizot, and P. Salati,  and support from the Israel-U.S. Binational Science Foundation, the Israeli Science Foundation. and the Joan and Robert Arnow Chair of Theoretical Astrophysics.  We thank  D. Kaplan for technical assistance with the numerical work, and M. Cirelli for assistance in using the PPPC data. NG acknowledges the support of the I-CORE Program of the Planning and Budgeting Committee and The Israel Science Foundation (grant 1829/12) and the Israel Space Agency (grant 3-10417).

\end{document}